\documentclass[12pt]{article}

\pdfoutput=1
\usepackage{amssymb,textcomp,amsmath,graphicx}
\DeclareMathAlphabet{\mathcal}{OMS}{cmsy}{m}{n}
\long\def\symbolfootnote[#1]#2{\begingroup
\def\thefootnote{\fnsymbol{footnote}}
\footnote[#1]{#2}\endgroup}

\newcommand{\be}{\begin{equation}} \newcommand{\ee}{\end{equation}}
\newcommand{\bea}{\begin{eqnarray}} \newcommand{\eea}{\end{eqnarray}}

\DeclareMathOperator{\erf}{erf}
\DeclareMathOperator{\erfc}{erfc}

\newcommand{\lp}{\left(}
\newcommand{\rp}{\right)}

\newcommand{\ie}{{\em i.e.~}}

\newcommand{\etal}{\emph{et\,al.~}}
\newcommand{\nn}{\nonumber}
\newcommand{\mpl}{M_{\rm P}}

\newcommand{\dphi}{{\dot\phi}}
\newcommand{\ddphi}{{\ddot\phi}}
\newcommand{\dchi}{{\dot\chi}}
\newcommand{\dPhi}{{\dot\Phi}}

\newcommand{\fNL}{{f_{\rm NL}^{\rm equil.}}}
\newcommand{\fNLz}{{f_{\rm NL,\,\zeta}^{\rm equil.}}}
\newcommand{\fNLt}{{f_{\rm NL,\,t}^{\rm equil.}}}

\newcommand{\ml}{{\mathcal{L}}}
\newcommand{\pz}{{\mathcal{P}_{\zeta}}}
\newcommand{\pzs}{{\mathcal{P}_{\zeta,\,{\rm src.}}}}
\newcommand{\pzv}{{\mathcal{P}_{\zeta,\,{\rm vac.}}}}

\newcommand{\pt}{{\mathcal{P}_{\rm t}}}
\newcommand{\ptv}{{\mathcal{P}_{\rm t,\,{\rm vac.}}}}
\newcommand{\pts}{{\mathcal{P}_{\rm t,\,{\rm src.}}}}

\newcommand{\p}{{\it{Planck 2015 }}}
\newcommand{\pp}{{particle production }}

\newcommand{\nG}{{non-Gaussian }}

\numberwithin{equation}{section}

\begin{document}
\begin{center}
\Large{\bf Primordial Black Holes Formation from Particle Production during Inflation}
\end{center}
\begin{center}
\large{Encieh Erfani$^{*}$}
\end{center}
\begin{center}
\textit{Department of Physics, Institute for Advanced Studies in Basic Sciences (IASBS),\\
No. 444, Prof. Yousef Sobouti Blvd., Zanjan, Iran\\
ICTP South American Institute for Fundamental Research,\\
IFT-UNESP, Rua Dr. Bento Teobaldo Ferraz 271, S\~{a}o Paulo, Brazil}
\end{center}

\date{}

\symbolfootnote[0]{$^{*}$erfani@iasbs.ac.ir}

\begin{abstract}

We study the possibility that \pp during inflation can source the required power spectrum for dark matter (DM) primordial black holes (PBH) formation.
We consider the scalar and the gauge quanta production in inflation models, where in the latter case, we focus in two sectors: inflaton coupled {\it i}) directly
and {\it ii}) gravitationally to a $U(1)$ gauge field.
We do not assume any specific potential for the inflaton field. Hence, in the gauge production case, in a model independent way we show that the non-production of
DM PBHs puts stronger upper bound on the \pp parameter. Our analysis show that this bound is more stringent than the bounds from the bispectrum and the tensor-to-scalar
ratio derived by gauge production in these models.
In the scenario where the inflaton field coupled to a scalar field, we put an upper bound on the amplitude of the generated scalar power spectrum by non-production of
PBHs.
As a by-product we also show that the required scalar power spectrum for PBHs formation is lower when the density perturbations are \nG in comparison to the
Gaussian density perturbations. 

\end{abstract}

\newpage

\section{Introduction}

\paragraph{}

Inflation is currently the standard paradigm for solving the cosmological puzzles of the standard big bang cosmology, such as homogeneity, isotropy and
flatness of the universe. In addition, the quantum fluctuation of the inflaton field explain the generation of all (classical) inhomogeneities that can
be seen in our universe, from the Cosmic Microwave Background (CMB) anisotropies to the Large Scale Structure (LSS) \cite{Inflation}.
In the simplest models, inflaton field slowly rolls down its potential, however, generically the inflaton should be expected to couple to some additional
degrees of freedom and a variety of different models have been proposed.\footnote{In general, the coupling of the inflaton to additional degrees of freedom
is necessary for reheating scenario after inflation \cite{reheating}.}

Recently, inflation models where production of some non-inflaton particles happens  {\it during} inflation via parametric resonance has received a lot
of attention\footnote{The phenomenon of resonant production of particles has received much attention in the past for successful (pre-)reheating {\it after}
inflation.} \cite{first, scalar-coupling, 0902.0615, 0908.4089, 0909.0751, 1305.3557, 1409.5799}. 
In this class of models, the inflaton field couples to another field directly or gravitationally and the coupled field can be massless or massive fermion
\cite{first}, scalar \cite{scalar-coupling, 0902.0615, 1511.06175} or gauge field \cite{9209238, 1011.1500, 1206.6117}. The production of light species occurs
during inflation at the expenses of the kinetic energy of the inflaton and slows down its motion; \ie resonant extraction of inflaton field energy
decreases $\dphi$, leading to an increase in the scalar power spectrum, $\pz\propto{H^4}/{\dphi^2}$. It was shown that the production of particles during inflation
provides a qualitatively new mechanism for generating cosmological perturbations \cite{first, 0908.4089, 0909.0751}. The nature of these fluctuations is usually
non-scale-invariant and \nG (NG) \cite{1011.1500, 1206.6117, 1102.4333, 1101.1525}.

A particular feature is that the power spectrum of these fluctuations can be very blue; this means that the amplitude of density fluctuations can be much
higher at the small length scales relevant for Primordial Black Holes (PBHs) formation \cite{Encieh1, Encieh2}.
To be consistent with cosmological observations such as the CMB and the LSS, these fluctuations should not dominate the primordial density perturbation
at large scales. However, depending on individual models, their significant contribution to the non-Gaussianity is still possible.
In this paper, we will show that these fluctuations with a blue spectrum in the very small scales, can seed the formation of Dark Matter (DM) PBHs,
therefore their non-production can constraint the \pp parameter which is a function of the parameter(s) of the inflation model.

Here, we will study the coupling of the inflaton to a massless scalar and a gauge field. 
In \cite{0902.0615}, a simple model where scalar $\chi$ particles are produced through a coupling $g^2\phi^2\chi^2$ to the inflaton, $\phi$ is considered.
In this model, the $\chi$ field whose mass depends on the inflaton becomes effectively massless as the inflaton rolls down its potential, therefore it
becomes energetically cheap to produce its quanta. In this case, instantaneous bursts of \pp leads to localized (bump-like) feature in the cosmological
perturbations. We will show that the non-production of PBHs with mass $\sim 10^{15}$ g -- as candidate for DM -- can put an upper bound on the amplitude
of these features.\footnote{In this case, the \pp can also leads to observable effects in the tensor power spectrum \cite{1109.0022}. However, in this
paper we are not interested in this effect.}

Another possibility is that particle production occurs continuously during inflation. This can happen when the inflaton, $\phi$ couples to a derivative
of some field such as a gauge field \cite{0908.4089, 9209238, 1102.4333}.
This is quite natural in the context of axion inflation \cite{natural-inflation} via the coupling of the pseudoscalar field to gauge fields,
$\alpha\phi\,F\tilde{F}$. This interaction leads to a continuous tachyonic production of gauge field fluctuations during inflation if the coupling,
$\alpha$ is large enough. 
It has been shown \cite{1011.1500, 1102.4333} that, these gauge fields through their inverse decay into inflaton perturbations 
-- \ie $\delta A + \delta A \rightarrow \delta\phi$ -- can affect all scalar primordial $N$-point functions. 
The effect on the power spectrum is a blue tilted contribution, which grows on smaller scales. Since the inverse decay produces NG inflaton perturbations,
the bispectrum is non-vanishing and peaks on equilateral configurations. In addition, gauge fields production can also leads to observable effects 
in the tensor power spectrum \cite{1206.6117, 1109.0022, 1110.3327}. The effect is negligible at the CMB scales but it could be detected by gravitational
wave (GW) interferometers such as advanced LIGO/Virgo \cite{LIGO-Virgo}\footnote{In \cite{1405.0346}, they claimed that particle production during
inflation can result in a GW detectable in the CMB scale.}.

Since the inverse decay of gauge fields into scalar perturbations leads to a blue tilt in the primordial scalar power spectrum it is natural to ask
what the constraint from the PBHs formation is.
As we will see in section~\ref{III}, at very small scales, the produced gauge quanta largely increases the curvature power spectrum which can reach
$\pz \sim\mathcal{O}(10^{-4})$ required for DM PBHs formation. Therefore, the non-production of such PBHs can put an upper bound on the scalar and
the tensor power spectra and the bispectrum.

The rest of the paper is organized as follows: In section~\ref{II}, we present a brief review of the Press-Schechter formalism \cite{Press-Schechter}
describing PBH formation by considering the new threshold derived in \cite{new-threshold} for their formation.
We will focus on non-evaporating PBHs as candidate for DM and we will discuss the choice of a Gaussian and a non-Gaussian probability distribution function (PDF)
for the perturbations as seed for their formation. And we will derive the required power spectrum for the production of such PBHs in both cases.
We will show that when the fluctuations are NG, the required spectral index (hence, the power spectrum) in the scale of DM PBHs is lower than 
the case when the perturbations are Gaussian. Therefore, the probability of DM PBHs formation is higher in the NG case.\\
In section~\ref{III}, we will review the \pp during inflation in different inflation models and we will estimate the scalar and the tensor power spectra and the bispectrum, and we will explore the possibility of DM PBHs formation in these models. 
In section~\ref{III.I}, we will review the main formulas and results of the axion inflation model in which the inflationary expansion is accompanied by the
gauge quanta production.
We will consider both cases when the inflaton couples directly and gravitationally to the gauge field. Our consideration differs from the consideration
carried out in the recent works \cite{1206.1685, 1212.1693, 1312.7435} in two respects. Firstly, we will not assume any specific potential for the inflaton
(or coupled) field. Secondly, we are only interested in DM PBHs; \ie the ones with mass $\gtrsim 10^{15}$ g.
We will show that in these cases the non-production of DM PBHs at the end of inflation, puts stronger upper bound on the particle production parameter, $\xi$. And this bound is more stringent than the bounds from the bispectrum and the tensor-to-scalar ratio derived by gauge production in these models.\\
We will also review a mechanism analogous to that of section~\ref{III.I}, where the gauge field is replaced by a massless scalar field in section~\ref{III.II}.
In a model independent way, we will put an upper bound on the amplitude of the (bump-like) feature of power spectrum -- generated by scalar production during inflation -- via the non-production of DM PBHs in the CMB observational range..\\ 
Finally, in section~\ref{IV} we present our conclusions.

Through this paper, we use the units $c=\hbar=1$.

\section{Constraints from Dark Matter Primordial Black Holes} \label{II}

\paragraph{}

Primordial black holes may have formed from primordial fluctuations in the early universe \cite{Hawking&Zeldovich}. 
Actually, there is no conclusive evidence for their existence and upper bounds on their abundance on various mass scales have been obtained by
various kinds of observations (for recent results, see \cite{bh-abundance}). These upper bounds can be translated into upper bounds on the power spectrum
of the scalar perturbation on comoving scales much  smaller than those relevant to the CMB; \ie they can be used to exclude inflationary models which
predict too many PBHs \cite{Encieh1, Encieh2}.\footnote{Recently, in \cite{1506.05228}, they claimed that PBHs can also constrain the amplitude of 
primordial tensor fluctuations.} On the other hand, according to the Hawking radiation \cite{Hawking-rad} the PBHs with mass  larger than $10^{15}$ g
which survive till present time could be candidate for DM.

The traditional treatment of PBH formation is based on the Press-Schechter formalism \cite{Press-Schechter} and in \cite{Encieh1}, 
this mechanism has been studied in detail for the formation of DM PBHs. Here, we briefly review this formalism by considering the new derived threshold
required for their formation. 
We will derive the required power spectrum for the production of DM PBHs in two cases: where the perturbations follow a {\it i}) Gaussian and a 
{\it ii}) non-Gaussian distribution. 

PBHs will form if at horizon re-entry, the amplitude of density perturbation passes a threshold value,
$\zeta_{\rm th} \equiv \lp\delta\rho/\rho\rp_{\rm th}$. The correct value of the threshold is quite uncertain. 
It was first provided by Carr \cite{Carr}, giving $\zeta_{\rm th}\simeq w$, where $w$ is the parameter of the equation of state, $p = w\rho$.
Recently, in \cite{new-threshold}, a new analytic formula for the threshold required for PBHs formation in the universe dominated by a
perfect fluid is derived. According to this formula, the amplitude of the density perturbations in the comoving hypersurfaces is given by
$\zeta_{\rm th}=\dfrac{3(1+w)}{5+3w}\delta_{\rm H}^{\rm UH}$, where $\delta_{\rm H}^{\rm UH} = \sin^2\lp\pi\sqrt{w}/(1+3w)\rp$
is the amplitude of the density perturbations at the horizon crossing time in the uniform Hubble slice. Therefore, $\zeta_{\rm th} \simeq 0.4135$
for a radiation fluid, $w = 1/3$.\footnote{Since the produced particles during inflation are massless, we can assume that the results of the
PBHs formation in the radiation dominated (RD) era after inflation is applicable in this case, too.}

\begin{figure}[ht]
\centering{\includegraphics[width=.95\textwidth]{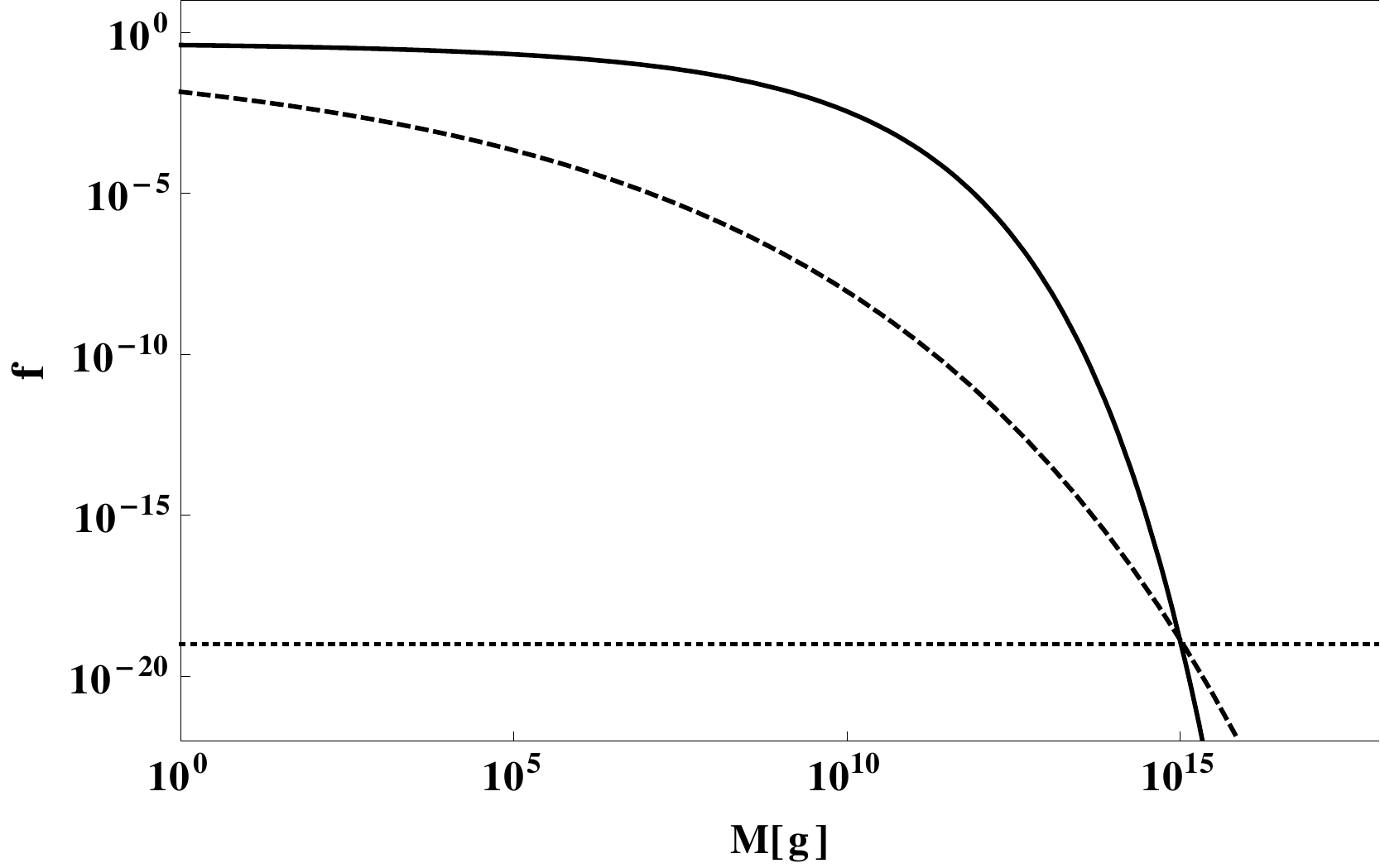}}
\caption{Fraction of the energy density of the universe collapsing into PBHs as a function of the PBH mass for the threshold $\zeta_{\rm th} = 0.4135$.
The solid (dashed) curve is for the Gaussian (non-Gaussian) probability distribution function. The horizontal dotted line indicates the abundance of
DM PBHs, $f \sim 10^{-19}$.}
\label{fig1}
\end{figure}

The fraction of the energy density of the universe in which the density fluctuation exceeds the threshold for PBH formation which will hence end up in PBHs
with mass $\geq \gamma M$\footnote{The PBH mass is a fraction $\gamma$ of the
horizon mass corresponding to the smoothing scale. A simple analytical calculation suggest $\gamma \simeq w^{3/2} \simeq 0.2$ during the radiation
era \cite{Carr}.}, is given by\footnote{In order to complete this calculation one needs to relate the mass, $M$ to the
comoving smoothing scale, $R$ in the RD era (see eq.~(2.10) in Ref.~\cite{Encieh1}). In this relation, the pre-factor changes from
$5.54\times10^{−24}$ to $5.19\times10^{−24}$ after the \p data \cite{Planck-XIII} due to the change in the matter-radiation equality redshift, $z_{\rm eq}$.}

\be \label{integral}
f(\geq M) = \gamma \int_{\zeta_{_{\rm th}}}^{\infty} P(\zeta;\,M(R))\,\text{d}\zeta\,. 
\ee
Here $P\lp\zeta;\,M(R)\rp$ is the PDF of the linear density field $\zeta$ smoothed on a scale $R$.

Hawking evaporation and present day gravitational effects constrain this fraction. For PBHs contributing to DM today, they saturate the DM relic density if
$f \simeq 5\times10^{-19}$ \cite{bh-abundance}.

For a derivation of the PBH constraints we need an expression for the PDF of the smoothed $\zeta$-field. The standard assumption is that it is Gaussian; \ie
\be
P_{\rm G}(\zeta;\,R) = \dfrac{1}{\sqrt{2\pi}\,\sigma_{\zeta}(R)} \exp\lp -\dfrac{\zeta^2}{2\sigma_{\zeta}^{2}(R)} \rp\,,
\ee
where $\sigma_\zeta^2(R) \equiv \langle \zeta^2 \rangle$ is the mass variance of $\zeta$.\\
Then
\be
f_{\rm G} = \frac{1}{2} \erfc \lp \zeta_{_{\rm th}}/\sqrt{2\sigma_{\zeta}^2(R)}\rp\,,
\ee
where $\erfc(x) \equiv 1-\erf(x)$ is the complementary error function. 

In some inflation models we will study in the next section, $\zeta$ follows a \nG distribution. Therefore, the simplest assumption which we
can use is that the $\zeta$-field is distributed as a square of some Gaussian field, $g$ \cite{NG-pbh}
\be
\zeta = g^2 - \langle g^2 \rangle\,.
\ee
Hence, the PDF of $\zeta$ takes the form
\be
P_{\rm NG}(\zeta;\,R) = \dfrac{1}{\sqrt{2\pi\lp\zeta+\sigma_{g}^2(R)\rp}\,\sigma_{g}(R)} \exp\lp -\dfrac {\zeta + \sigma_{g}^2(R)}{2\sigma_{g}^{2}(R)} \rp\,.
\ee
Note that here, $\sigma_\zeta^2(R) = 2 \langle g^2 \rangle^2 \equiv 2 \lp\sigma_g^2(R)\rp^2$. Therefore, the fraction of space that can collapse to
form black holes in a case $\zeta$ dose not follow a Gaussian distribution is given by
\be
f_{\rm NG} = \erfc \lp \sqrt{\zeta_{_{\rm th}} + \sigma_{g}^2(R)}/\sqrt{2\sigma_{g}^2(R)} \rp\,.
\ee

Following the procedure in Ref.~\cite{Encieh1}, we show the final result in Fig.~\ref{fig1}. Here we have fixed $\zeta_{\rm th} = 0.4135$,
and show the results for Gaussian (solid curve) and \nG (dashed curve) probability distribution function.
We found that for the formation of DM PBHs from density perturbations with a power-law spectrum, the spectral index at scale of PBHs formation
($k_{\rm PBH}$)\footnote{The relevant scale for the formation of DM PBHs is $k_{\rm PBH} = 1.52\times10^{15}$ Mpc$^{-1}$ \cite{Encieh1}.} 
should be at least $1.418$\footnote{Note that this value is larger than the value found in Ref.~\cite{Encieh1}. This is due to the fact that here
the larger threshold value used for the amplitude of the density perturbations at the horizon crossing.} $(1.322)$ for Gaussian (non-Gaussian) PDF.
Therefore, the DM PBHs formation can only happen if the spectral index increases significantly between the scales probed by the CMB and the relevant
scales for DM PBHs. With $n_{\rm s}(k_{\rm PBH}) \simeq 1.418\,(1.322)$, the power spectrum at the scale of DM PBHs formation should
be $\pz(k_{\rm PBH}) \simeq 2\times10^{-2}\,(4\times10^{-4})$, which is $\sim 10^{7}(10^{5})$ times higher than the value of the power spectrum
in the observed CMB scales. Note that for the production of PBHs in the \nG case, one needs a lower power spectrum in comparison to the Gaussian case.

\section{Particle Production during Inflation Models} \label{III}

\paragraph{}

In this section, we discuss the fluctuations arised during inflation via the direct or gravitational coupling of the inflaton to another field.
We study different inflation models and explore the possibility of DM PBHs formation from these fluctuations.

We consider a model where the slow-rolling inflaton, $\phi$ coupled to a field $\chi$\footnote{Through this paper we will
denote the inflaton field and the coupled field by $\phi$ and $\chi$, respectively.}. 
The relevant Lagrangian density is given by
\be
\ml (\phi,\,\chi) = -\dfrac{1}{2} \partial_{\mu}\phi\,\partial^{\mu}\phi - V (\phi) 
-\dfrac{1}{2} \partial_{\mu}\chi\,\partial^{\mu}\chi - U(\chi) + \ml_{\rm int}(\phi,\,\chi)\,, 
\ee
where $V(\phi)$ is the potential supporting inflation, $U(\chi)$ is the potential of $\chi$ field and the fields interact via $\mathcal{L}_{\rm int}$ 
term.\footnote{We will assume that only the inflaton potential is dominant during the inflation and we will neglect the potential of the coupled field, $\chi$.}

The equations of motion for the inflaton field is given by
\be 
H^2 = \dfrac{1}{3\mpl^2} \lp\frac{1}{2}\dphi^2 + V(\phi) + \rho_\chi\rp\,,
\ee
\be \label{eom}
\ddphi + 3H\,\dphi + V^\prime(\phi) = \dfrac{\partial\ml_{\rm int}}{\partial\phi}\,,
\ee
where $\rho_\chi$ is the energy density of $\chi$ particles, and the prime and the dot denote differentiating with respect to $\phi$ and time,
respectively.\footnote{$\mpl = 1/\sqrt{8\pi G} \simeq 2.4\times10^{18}$ GeV is the reduced Planck mass.} The right hand side of the eq.~\eqref{eom}, 
is the backreaction of the interaction to the inflaton mean field which arises due to copious production of $\chi$ quanta during inflation.

In addition, the inflaton fluctuations satisfy
\be \label{fluc-eq}
\ddot{\delta\phi} + 3H\,\dot{\delta\phi} - \dfrac{\bigtriangledown^2}{a^2} \delta\phi + V^{\prime\prime}(\phi)\,\delta\phi = 
\delta \lp\dfrac{\partial\ml_{\rm int}}{\partial\phi}\rp\,.
\ee
The solution of the above equation without the interaction term gives rise to the primordial density fluctuations and we denote their power spectrum
by $\pzv$. However, the right hand side of eq.~\eqref{fluc-eq} acts as a source for generating additional fluctuations of $\phi$ and the corresponding 
power spectrum is denoted by $\pzs$. Hence, the total power spectrum is given by the contributions from both fluctuations as
\be
\pz(k) = \pzv(k) + \pzs(k)\,.
\ee

From the observational perspective, the key quantities which characterize any model of inflation are the spectrum of scalar and tensor perturbations,
$\pz$ and $\pt$, respectively, along with the bispectrum of scalar perturbations, $B_{\zeta}$ that encodes the leading departures from Gaussian statistics.
Therefore, before studying different inflationary models, it is worth defining these cosmological observables and their current constraints.

The power spectrum of the curvature perturbations of inflaton field is given by
\be
\pzv = \dfrac{1}{12\pi^2\mpl^6}\dfrac{V^3(\phi)}{{V^\prime}^2(\phi)} = \dfrac{1}{24\pi^2\mpl^4}\dfrac{V(\phi)}{\epsilon} \simeq \dfrac{H^2}{4\pi^2\dphi^2}\,,
\ee
where $\epsilon \equiv \dfrac{\mpl^2}{2}\lp\dfrac{{V^\prime}(\phi)}{V(\phi)}\rp^2$ is the first slow-roll parameter.\\
On the other hand, the simplest assumption for the power spectrum is a scale-free power-law
\be
\pzv(k) = \pzv(k_0) \lp\dfrac{k}{k_0}\rp^{n_{\rm s}(k)-1}\,,
\ee
where the scale dependence of $\pzv$ is parameterized by the spectral index
\be
n_{\rm s}(k_0)-1 \equiv \dfrac{d\ln\pzv(k)}{d\ln k}\,.
\ee

During inflation, GW fluctuations are generated by quantum fluctuations of the tensor part of the metric. The power spectrum
of these perturbation from vacuum fluctuations during inflation is given by
\be
\ptv(k) = \dfrac{2}{\pi^2}\lp\dfrac{H}{\mpl}\rp^2 \lp\dfrac{k}{k_0}\rp^{n_{\rm t}}\,,
\ee
where $n_{\rm t} = -2\epsilon$.\\
Additional sources of GW may arise due to the coupling of the inflaton to another field. And since the power of the vacuum and the sourced mode are incoherent,
they add up \cite{1206.6117}
\be
\pt(k) = \ptv(k) + \pts(k)\,.
\ee
On the other hand, the tensor contribution to the power spectrum is parameterized by the tensor-to-scalar ratio, $r$
\be
r = \dfrac{\pt(k)}{\pz(k)}\,,
\ee
and $r = 16\epsilon$ if we consider the primordial scalar and tensor power spectra.

Non-Gaussian statistics, such as the bispectrum, provide a powerful tool to discriminate between the inflation models and, $f_{\rm NL}$ characterizes
its amplitude
\be
B_{\zeta}(k_1,\,k_2,\,k_3) = f_{\rm NL}F(k_1,\,k_2,\,k_3)\,,
\ee
where the function $F(k_1,\,k_2,\,k_3)$ describes the shape of the bispectrum.

Now let us review the observational bounds on inflation parameters with the {\it Planck} ``TT,\,TE,\,
EE+lowP'' data \cite{Planck-XX, Planck-XVII} at the $68\,\%$ confidence level. Note that \p full mission data in temperature and polarization  are quoted at the 
pivot scale $k_0 = 0.05$ Mpc$^{-1}$.\\
For the base $\Lambda$CDM model, the constraints on the power spectrum, $\pzv(k_0)$
and the scalar spectral index, $n_{\rm s}$ are \cite{Planck-XX}
\bea
\ln(10^{10}\pzv(k_0)) &=& 3.094 \pm 0.034\,,\label{pzv}\\
n_{\rm s} &=& 0.9645 \pm 0.0049\,.
\eea

The constrain on the tensor-to-scalar ratio inferred from the \p data for the $\Lambda$CDM+$r$ model is\footnote{Note that the tensor-to-scalar ratio is reported
at $k_0 = 0.002$ Mpc$^{-1}$.}\cite{Planck-XX}
\be
r_{0.002} < 0.10 \qquad (95\,\%\,{\rm CL})\,.
\ee
Since in the inflation models we will study in the following section, the inverse decay type NG is present, we only report the observational constraint obtained in \p 
{\it XVII} for this type \cite{Planck-XVII}
\be \label{NL-Obs}
f_{\rm NL} = 22.7\pm25.5\,.
\ee

In the rest of the paper, we study the gauge field production in two cases: {\it i}) the inflaton field coupled directly to the gauge field
(section~\ref{III.I.I}) and {\it ii}) the inflaton field coupled gravitationally to the gauge field (section~\ref{III.I.II}). 
In section~\ref{III.II} we consider the scalar particles production via the coupling of the inflaton to a scalar field. 

\subsection{Gauge Production} \label{III.I}

\paragraph{} 

In this section we review the mechanism by which gauge fields are resonantly enhanced when a pseudoscalar field couples to a massless $U(1)$ gauge field. 
When the coupling is direct, the pseudoscalar field will play the roll of the inflaton \cite{1011.1500}.
It is also possible that the inflaton field couples gravitationally to a pseudoscalar field which is coupled directly to the gauge field \cite{1206.6117}. 

We can write a generic interaction Lagrangian which is compatible with gauge symmetry as follows \cite{1110.3327}

\be \label{interaction}
\ml_{\rm int} = - \dfrac{1}{4}F_{\mu\nu}F^{\mu\nu} - \dfrac{\alpha}{4f}\Phi\,F_{\mu\nu} \tilde{F}^{\mu\nu}\,,
\ee
where $\Phi$ could be the inflaton or any other pseudoscalar field and $F_{\mu\nu} = \partial_\mu A_{\nu} - \partial_{\nu}A_{\mu}$ is the field
strength associated with gauge field, $A_\mu$\footnote{For simplicity this gauge field is not the Standard Model one.} and
$\tilde{F}_{\mu\nu}=\epsilon_{\mu\nu\alpha\beta}F^{\alpha\beta}/2$ is its dual. The strength of the interaction is controlled by the dimensionless parameter,
$\alpha$ and the decay constant, $f$.  

In this type of scenario, gauge fields can be generated during inflation and has a rich and interesting phenomenology both for scalar and tensor
primordial fluctuations \cite{1011.1500, 1101.1525}. Also in this case the inverse decay of gauge field perturbations leads to a large
equilateral\footnote{When a pseudoscalar couples to a vector, non-Gaussianity can feature equilateral configuration because in this model, the source at any 
moment is dominated by modes with wavelength comparable to the horizon at that moment. This generates mostly correlations between scalar perturbations
of comparable size \cite{1011.1500}.} contribution to the bispectrum \cite{1011.1500, 1102.4333}.

It was shown \cite{0908.4089, 1011.1500, 1102.4333} that the inverse decay of the gauge field source additional scalar fluctuations to the inflaton field
which is given by
\be \label{scalar}
\pzs = \gamma_{\rm s}\,\epsilon^2\,\mathcal{P}_{\zeta,\,{\rm vac.}}^{2} X^2\,,
\ee
where $X \equiv \dfrac{e^{2\pi\xi}}{\xi^3}$ and $\xi\equiv \dfrac{\alpha}{2fH}\dPhi$, is a dimensionless parameter which characterizes
the strength of the inverse decay effects. It is clear that when $\xi > 0$, the phase space of produced fluctuations experiences a significant 
exponential enhancement, $e^{\pi\xi} \gg 1$, due to tachyonic instability near horizon crossing.

The produced gauge quanta also source tensor metric perturbations \cite{1011.1500} and it was pointed out that these modes are chiral \cite{1101.1525}
\be \label{tensor1}
\mathcal{P}_{\rm t,\,{\rm src.}}^{\pm} = \gamma_{\rm t}^{\pm}\,\mathcal{P}_{\rm t,\,{\rm vac.}}^{2}X^2\,,
\ee
and, as a consequence of the violation of the parity, the left and right-handed tensor modes have different amplitude of the spectra.

The tensor-to-scalar ratio is as follows
\be \label{r}
r = \dfrac{\pt^{+}+\pt^{-}}{\pz}\,.
\ee

The inverse decay contribution to the primordial cosmological fluctuations is highly NG and is given by
\be \label{NG}
\fNL = \gamma_{_{\rm NG}} \epsilon^3\,\pzv X^3\,.
\ee

Note that in eqs.~\eqref{scalar}, \eqref{tensor1} and \eqref{NG}, the coefficients $\gamma_{\rm s}$, $\gamma_{\rm t}$ and $\gamma_{_{\rm NG}}$ are dimensionless
parameters and their precise value needs to be computed in inflation models that we will study in the following sections.

It is clear that in inflation scenarios associated with a pseudoscalar coupling to a $U(1)$ gauge field, the observables (power spectrum, NG and tensor-to-scalar
ratio) depend only on the model-dependent quantities $\xi$ and $\epsilon$, which in turn depend on the inflation potential and the dynamics of the coupled field.

In the following we review the inflation models when a pseudoscalar field plays the roll of inflaton or not and we study the possibility of DM PBHs formation 
in these models. We check whether the bounds from the non-production of such PBHs is stronger than the bounds from NG and tensor perturbations or not. Note that we leave
$V(\phi)$ arbitrary, except to suppose that it is sufficiently flat to support $N\gtrsim 60$ $e$--foldings of inflation.

\subsubsection{Model I: Inflaton Coupled Directly to the Gauge Field} \label{III.I.I}

\paragraph{}

We consider a simple theory of a PNGB inflaton interacting directly with a massless $U(1)$ gauge field via the interaction \eqref{interaction} \cite{1011.1500}.

It was shown \cite{1011.1500, 1102.4333}, in this model $\gamma_{\rm s} \simeq 7.5\times10^{-5}$ (see eq.~\eqref{interaction}).
Therefore the total scalar power spectrum is given by
\be \label{scalar-I}
\pz = \pzv \lp 1 + 7.5\times10^{-5}\,\epsilon^2\,\pzv X^2 \rp\,.
\ee

\begin{figure}[ht]
\centering{\includegraphics[width=1\textwidth]{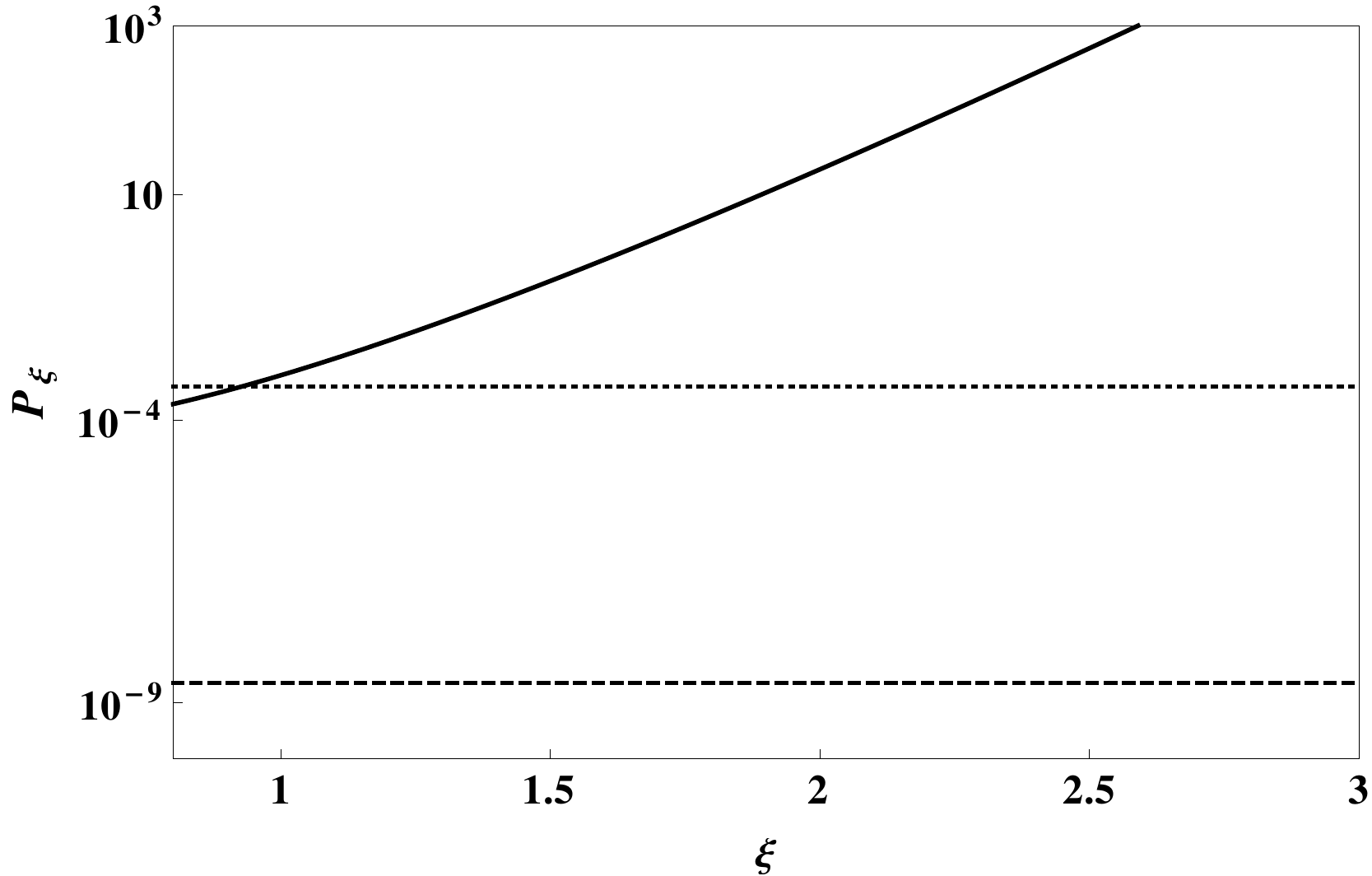}}
\caption{The total curvature power spectrum, $\pz$ as a function of \pp parameter, $\xi$. The DM PBHs bound is the dotted line and the dashed line shows the
constraint on the power spectrum from the CMB \cite{Planck-XX}.}
\label{gauge1}
\end{figure}

For tensor perturbations, it was shown \cite{1101.1525} that $\gamma_{\rm t}^{+} = 4.3\times10^{-7}$ and $\gamma_{\rm t}^{-} = 9.2\times10^{-10}$. 
Hence, the overall power spectra of the helicity-$\pm$ components of the GWs generated by this mechanism are as follows 
\bea \label{tensor-I}
\mathcal{P}_{\rm t,\,{\rm src.}}^{+} &=& 4.3\times10^{-7}\,\mathcal{P}_{\rm t,\,{\rm vac.}}^2X^2  
= 5.5\times10^{-5}\,\epsilon^{2}\,\mathcal{P}_{\zeta,\,{\rm vac.}}^{2}X^{2}\,,\nn\\
\mathcal{P}_{\rm t,\,{\rm src.}}^{-} &=& 9.2\times10^{-10}\,\mathcal{P}_{\rm t,\,{\rm vac.}}^2X^2
= 1.2\times10^{-7}\,\epsilon^{2}\,\mathcal{P}_{\zeta,\,{\rm vac.}}^{2}X^{2}\,,
\eea
and the tensor-to-scalar ratio is given by
\be \label{r-I}
r = 16\epsilon\,\dfrac{1 + 2.2\times10^{-7}\,\ptv\,X^2}{1 + 7.5\times10^{-5}\,\epsilon^2\,\mathcal{P}_{\zeta,\,{\rm vac.}} X^2}\,.
\ee

The inverse decay contribution to the primordial cosmological fluctuations lead to equilateral non-Gaussianity in the CMB which is given by \cite{1011.1500}
\be \label{NGs-I}
\fNLz \approx 4.4\times10^{10}\,\epsilon^3\,\mathcal{P}_{\zeta,\,{\rm vac.}}^3\,X^3\,.
\ee

We can plot these results for $\pz$, $r$ and $\fNLz$ as a function of $\xi$, for various representative choices of the slow-roll parameter, $\epsilon$.
However, since we are interested in \pp near the end of inflation we fix $\epsilon = 1$.  The total scalar power spectrum is plotted in Fig.~\ref{gauge1}
by using eq.~\eqref{scalar-I}. It is clear from the figure that when $\xi \sim 0.93$, the power spectrum can reach $4\times10^{-4}$ which is required value
for long--lived PBHs formation. 

\begin{figure}[ht]
\centering{
\includegraphics[width=0.48\textwidth]{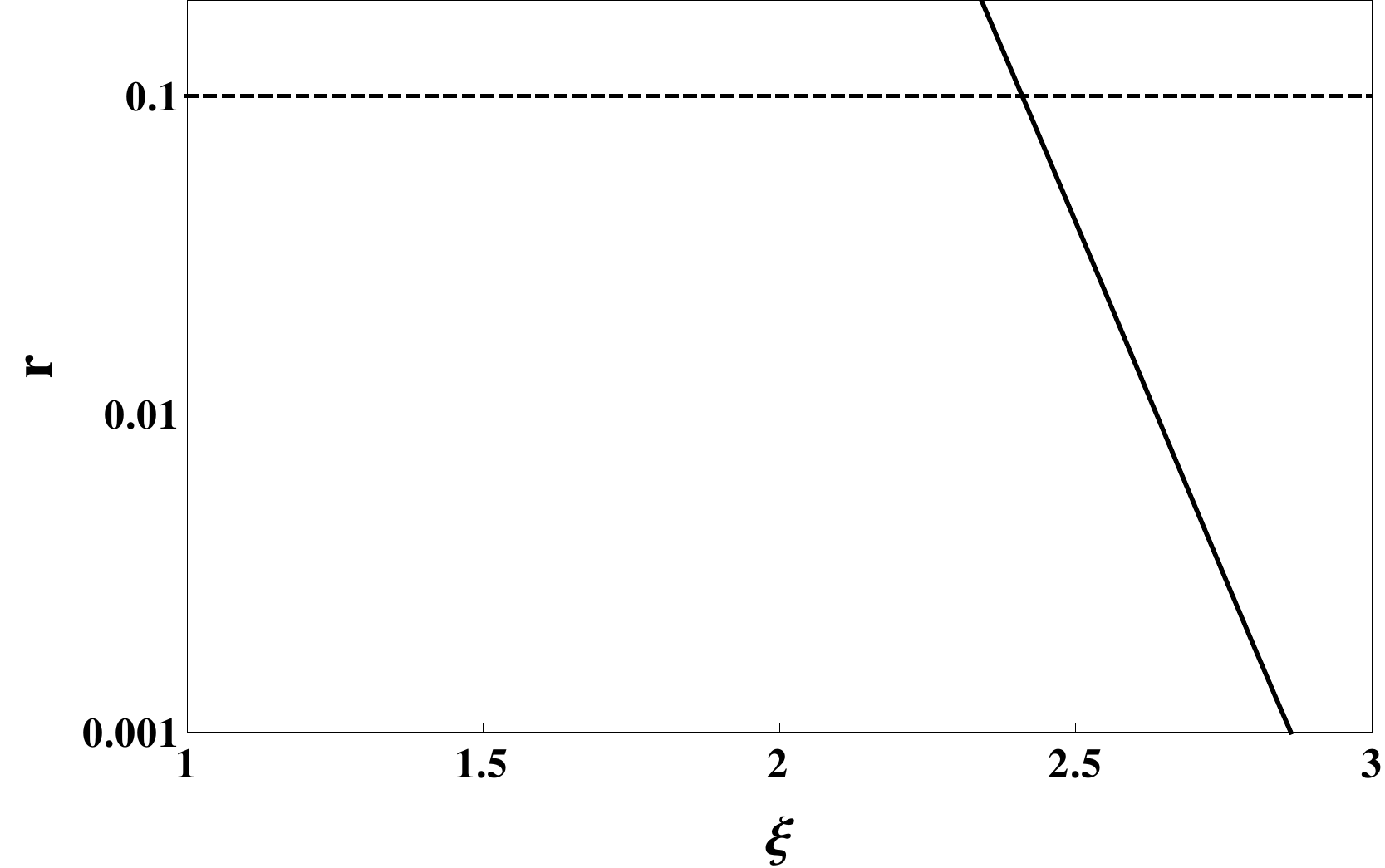}
\hspace{0.1cm}
\includegraphics[width=0.48\textwidth]{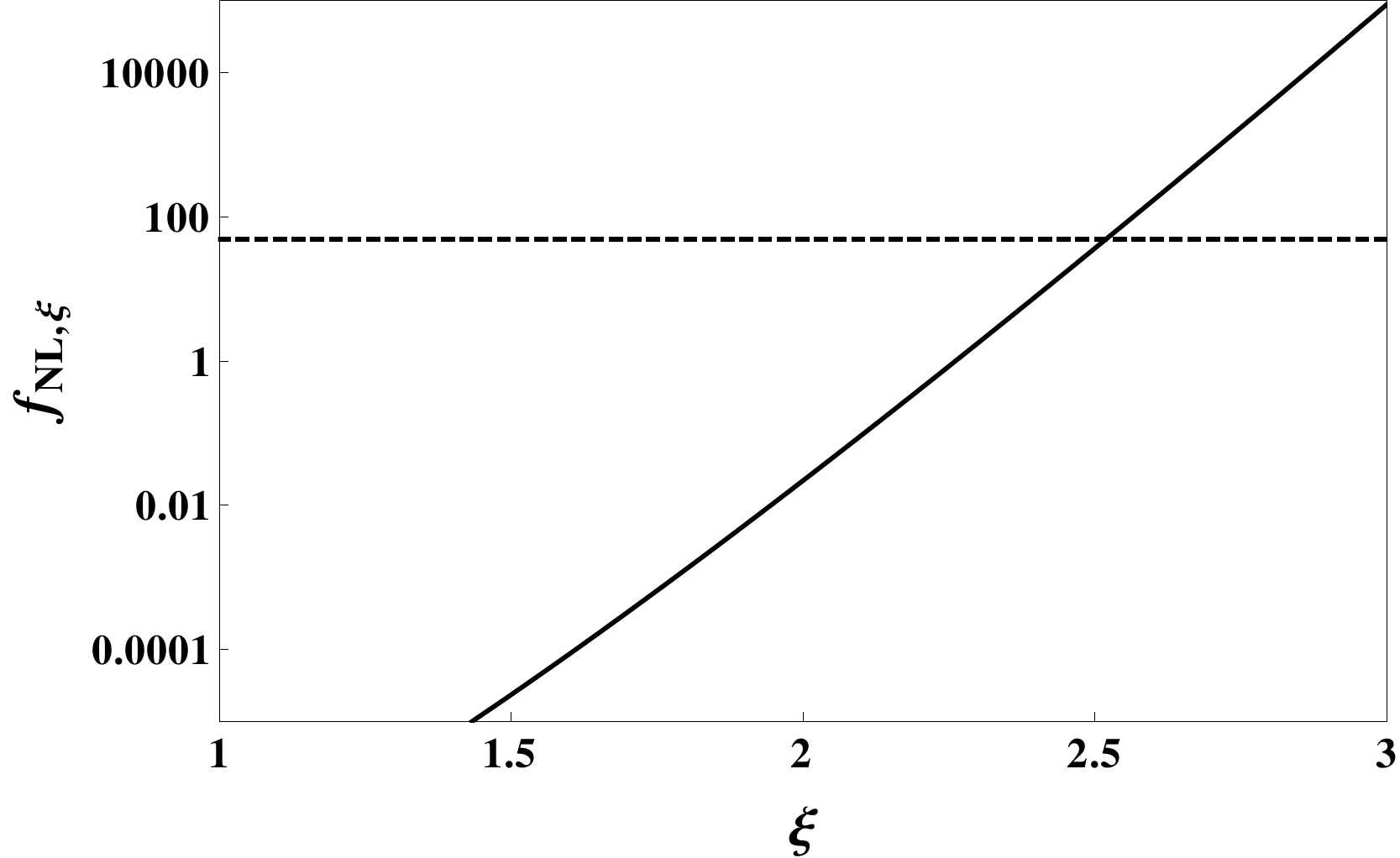}}
\caption{Predicted values for the tensor-to-scalar ratio, $r$ (left panel) and the equilateral $\fNLz$ parameter (right plot) as a function of $\xi$ near the
end of inflation. In left panel, the dashed line is \p \cite{Planck-XX} upper bound $r < 0.10$. The dashed line in the right plot is the
$68\,\%$ CL upper bound for $\fNLz$ \cite{Planck-XVII}.}
\label{gauge2}
\end{figure}

From the left plot of Fig.~\ref{gauge2} it is clear that the sourced GWs dominate over the vacuum ones for $\xi \gtrsim 2.41$ when $r=0.10$.

Using \eqref{NGs-I}, we see that the $2\,\sigma$ \p limit \eqref{NL-Obs} would enforce $\xi \lesssim 2.52$ (see right panel of Fig.~\ref{gauge2}). 

As a result, our analysis show that the most stringent constraints on $\xi$ is derived from the non-production of DM PBHs at the end of inflation and the bounds from the bispectrum and the tensor-to-scalar ratio are weaker. However, the bound from the tensor-to-scalar ratio is slightly more stringent than the NG bound. It is worth mentioning that we can also consider the PBHs formation during inflation (\ie $\epsilon < 1$), however the upper bounds on $\xi$ will be weaker.

\subsubsection{Model II: Inflaton Coupled Gravitationally to the Gauge Field} \label{III.I.II}

\paragraph{}

Above we discussed the possibility of DM PBHs where the gauge quanta are produced by a direct coupling between the inflaton and the gauge field. 
In this section we study the gauge production in a ``hidden sector'' where the vector and the pseudoscalar are only gravitationally coupled to the inflaton.
In this case a rolling pseudoscalar sources gauge filed fluctuations continuously. 

In the following we consider the system introduced in \cite{1206.6117}, in which a pseudoscalar field, $\chi$ couples to a $U(1)$ gauge filed
\be
\ml = - \underbrace{\dfrac{1}{2} \partial_{\mu}\phi\,\partial^{\mu}\phi - V(\phi)}_{\rm inflaton\,\,sector} 
- \underbrace{\dfrac{1}{2} \partial_{\mu}\chi\,\partial^{\mu}\chi - U(\chi) - \dfrac{1}{4}F_{\mu\nu}F^{\mu\nu} 
- \dfrac{\alpha}{4f}\chi F_{\mu\nu} \tilde{F}^{\mu\nu}}_{\rm hidden\,\,sector}\,.
\ee

Here the assumption is that the $\chi-A_\mu$ sector has a negligible energy density with respect to the inflationary sector, and the inflaton sector and
the hidden sector are coupled only gravitationally. 
In this case, gravitational coupling will transmit the effects of particle production in the hidden sector to the inflaton sector, and as in
section~\ref{III.I.I}, the production of gauge field fluctuations can provide a new source of scalar perturbations, complementary to the usual
quantum vacuum fluctuations of the scalar part of the metric. The total spectrum of the scalar perturbations was found in \cite{1206.6117} to be 
\be \label{scalar-II}
\pz \approx \pzv \lp 1 + 2.5\times10^{-6}\,\epsilon^2\,\mathcal{P}_{\zeta,\,{\rm vac.}} X^2 \rp\,,
\ee
therefore $\gamma_{\rm s} \simeq 2.5\times10^{-6}$. Note that here, $\xi \equiv \dfrac{\alpha}{2fH}\dchi$. 

Even if the vectors are not directly coupled to the inflaton, they generate tensor perturbations \cite{1206.6117} which are chiral 
\bea \label{tensor-II}
\mathcal{P}_{\rm t,\,{\rm src.}}^{+} &=& 2.13\times10^{-6}\,\mathcal{P}_{\rm t,\,{\rm vac.}}^2 X^2  
= 5.41\times10^{-4}\,\epsilon^2\,\mathcal{P}_{\zeta,\,{\rm vac.}}^{2}X^{2}\,,\nn\\
\mathcal{P}_{\rm t,\,{\rm src.}}^{-} &=& 4.44\times10^{-7}\,\mathcal{P}_{\rm t,\,{\rm vac.}}^2 X^2
= 1.14\times10^{-6}\,\epsilon^2\,\mathcal{P}_{\zeta,\,{\rm vac.}}^{2}X^{2}\,.
\eea

From eqs.~\eqref{scalar-II} and \eqref{tensor-II} we can write the tensor-to-scalar ratio as
\be \label{r-II}
r = 16\epsilon\,\dfrac{1 + 1.3\times10^{-6}\,\ptv\,X^2}{1 + 2.5\times10^{-6}\,\epsilon^2\,\mathcal{P}_{\zeta,\,{\rm vac.}} X^2}\,.
\ee

The scalar bispectrum is given by \cite{1206.6117}
\be \label{NGs-II}
\fNLz \approx 2.31\times10^{8}\,\epsilon^3\,\mathcal{P}_{\zeta,\,{\rm vac.}}^3\,X^3\,.
\ee
On the other hand, in this model, the tensor bispectrum also generates via the gravitational interaction with the gauge field, and its amplitude
is estimated as \cite{1307.7077} 
\be \label{NGt-II}
\fNLt \approx 1.1\times10^{12}\,\epsilon^3\,\mathcal{P}_{\zeta,\,{\rm vac.}}^3\,X^3\,.
\ee
It is clear that the non-Gaussianity in the tensor perturbations is higher than the scalar perturbations.
 
In Table~\ref{tab1}, we report the obtained upper bounds on the \pp parameter, $\xi$ from the non-production of DM PBHs at the end of inflation and the observational bounds on the tensor-to-scalar ratio and the scalar and the tensor non-Gaussianity.  

\begin{table}[h!]
\centering
\begin{tabular}{ | c | c | }
  \cline{2-2}
  \multicolumn{1}{c|}{}  & $\xi$    \\ \cline{1-2}
  $\pz$                  & $0.93$   \\ \cline{1-2}
  $r$                    & $2.41$   \\ \cline{1-2}
  $\fNLz$                & $2.86$   \\ \cline{1-2}
  $\fNLt$                & $2.31$   \\ \cline{1-2}
\end{tabular}
\caption{The obtained upper bounds on parameter $\xi$ by different observations (see the main text for details.).}
\label{tab1}
\end{table}

In this model also the bound on \pp parameter from the non-production of DM PBH is stronger than the ones derived from the (scalar and tensor) bispectrum 
and the tensor-to-scalar ratio. Note that the bound on $\xi$ from the tensor non-Gaussianity is stringent than the tensor-to-scalar ratio in this model.

\subsection{Scalar Production} \label{III.II}

\paragraph{} 

In this final section we consider a very simple model where a slow-rolling inflaton, $\phi$ interacts with a massless scalar field, $\chi$ via the
coupling \cite{0909.0751}
\be \label{scalar-model}
\ml_{\rm int}(\phi,\,\chi) = -\frac{g^2}{2}\lp\phi - \phi_0\rp^2\chi^2\,,
\ee
where $g$ is a coupling constant and $\phi_0$ is a constant field value. As $\phi(t)$ rolls down its potential, at the moment when
$\phi = \phi_0$, the effective mass of $\chi$ -- \ie $m_\chi(t) \equiv g\,(\phi(t)-\phi_0)$ -- becomes zero and its quanta can be copiously produced.
This burst of particle production extracts energy from the condensate $\phi(t)$ and temporarily slows down the motion of the inflaton field.\footnote{It is
also possible that several instances of \pp happens which gives rise to a new inflationary mechanism, called trapped inflation \cite{trapped-inflation}.
In this case multiple bursts of \pp leads many localized features in the power spectrum.
And since the production of particles occurs several times per $e$--folding of inflation, the inflaton slows down, therefore inflation can occur
even on a (relatively) steep potential. For more details see \cite{1109.0022, 1101.4493}.}
It was shown \cite{0902.0615}, the resonant extraction of even a small fraction of the energy of the inflaton field can alter the classical motion of the 
inflaton and generate a bump-like feature in the primordial power spectrum which in the model \eqref{scalar-model} can be fit with a following function
\be
\pzs(k) \sim k^3 e^{-\frac{\pi}{2}\lp\frac{k}{k_*}\rp^2}\,.
\ee
The location of the peak in wavenumber, $k_*$ depends on where the mass of the $\chi$ field passes through zero.\footnote{We should mention that the bump-like feature in the power spectrum can also produce in hybrid inflation models due to waterfall field transition. Therefore, PBHs can form in this case, too. For recent work in this subject, see \cite{1501.07565}.}

It is straight forward to generalize this model to allow for multiple bursts of particle production during inflation \cite{0902.0615}
\be \label{scalar-gen}
\ml_{\rm int}(\phi_i,\,\chi_i) = -\sum_{i=0}^{n}\frac{g_i^2}{2}(\phi-\phi_i)^2\chi_i^2\,.
\ee
In this case the contributed power spectrum from the burst of the produced $\chi_i$ particles is given by \cite{0902.0615}
\be
\pzs(k) = \sum_{i=1}^{n} A_i \lp\dfrac{\pi e}{3}\rp^{3/2} \lp\dfrac{k}{k_i}\rp^3 e^{-\frac{\pi}{2}\lp\frac{k}{k_i}\rp^2}\,,
\ee
where the constants, $A_i$ are the amplitudes of the power in the $i$-th bump which depend on the couplings, $g_i^2$ and, $k_i$ defines 
the location of features.

\begin{figure}[h!]
\centering{\includegraphics[width=1\textwidth]{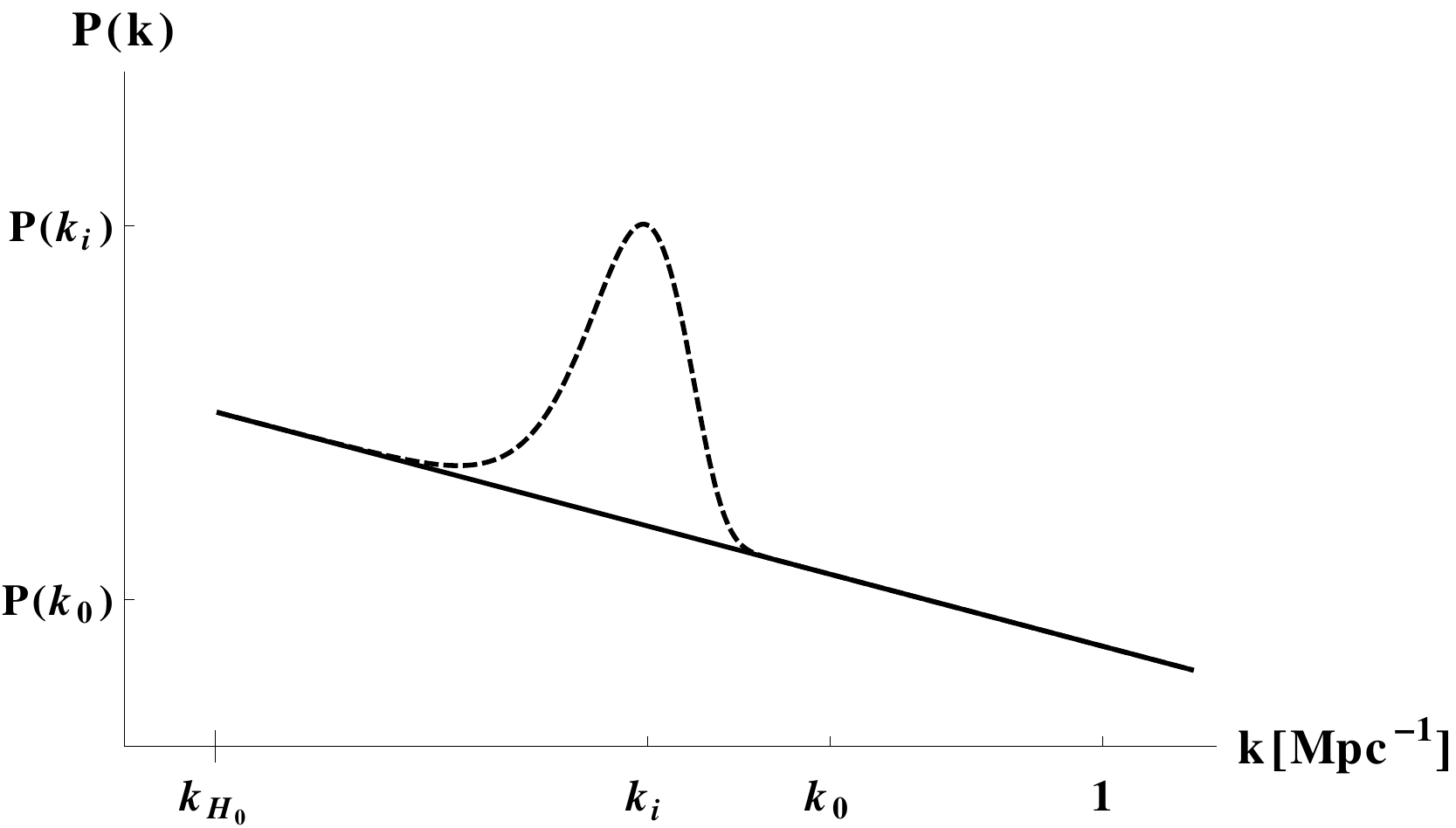}}
\caption{Power spectrum of curvature perturbations in the absence (solid curve) and the presence (dashed curve) of resonant particle creation in the
{\it Planck} observational range; \ie $k_{H_0} \lesssim k \lesssim 1$ Mpc$^{-1}$. Resonant particle production produces a peak at $k_i = 0.01$ Mpc$^{-1}$.}
\label{fig-scalar}
\end{figure}

In this model, the sample bump in the power spectrum is illustrated in Fig.~\ref{fig-scalar} and we take the bump to be located at $k_i = 0.01$ Mpc$^{-1}$.

As mentioned in section~\ref{II}, if the power spectrum is $\pz \sim \mathcal{O}(10^{-4})$, DM PBHs will be copiously produced in a scale corresponding to their masses. Here, to avoid their production in the CMB observational range (\ie $k_{H_0} \lesssim k \lesssim 1$ Mpc$^{-1}$), we can put an upper bound on the amplitude of the generated feature (and hence the couplings, $g_i$) due to the particle production; \ie $A_i \lesssim 4\times10^{-4}$.
Note that this bound is weaker than the one in \cite{0909.0751}, however, our result is completely model independent since we do not assume any specific potential for the inflaton field (\ie the energy scale of inflation), therefore we can not put any bound on the abundance of produced (mono-chromatic) PBHs.

\section{Summary and Conclusions} \label{IV}

\paragraph{} 

In this work we reviewed the new class of inflation models where the inflaton field couples to an other field. The inclusion of this coupling is that
it can produce the quanta of the coupled field. In this article we have studied two realizations of this scenario: where the coupled field is a vector
or a scalar field. In the first case we studied the direct and gravitational coupling of the inflaton to the $U(1)$ gauge field. In this model, inflation 
is accompanied by the gauge quanta production and a strong rise of the curvature and tensor power spectrum amplitude at small scales along with equilateral
non-Gaussianity is predicted.
However, we have estimated that towards the end of inflation, \ie on scales that are too small to be observed in the CMB, the power spectrum grows so much
that the model may be ruled out because of the overproduction of (dark matter) primordial black holes. 
Our consideration was different from the studies in recent works \cite{1206.1685, 1212.1693, 1312.7435} in two respects. Firstly, we did not assume any specific
potential for the inflaton (or coupled) field. Secondly, we were only interested in PBHs with mass larger than $10^{15}$ g as candidate for DM.
Since in the mentioned inflation models, the perturbations are highly non-Gaussian, we study the DM PBHs formation when the PDF of perturbations is non-Gaussian.
Our analysis in section~\ref{II} showed that the power spectrum at the scale of DM PBHs formation should be $\pz(k_{\rm PBH}) \simeq 4\times10^{-4}$,
which is two orders of magnitude smaller than the required scalar power spectrum when the perturbations are Gaussian. This means that the 
production probability of the non-evaporating PBHs when the perturbations are \nG is higher than the Gaussian case.

The main results when the inflaton coupled directly to the gauge field were shown in Figs.~\ref{gauge1} and \ref{gauge2}. From Fig.~\ref{gauge1} it was clear that 
due to tachyonic instability of gauge field, an amplitude of the scalar power spectrum, $\pz$ can reach $\sim10^{-4}$, which is required value for DM PBHs 
formation. Hence, we could put an upper bound on the \pp parameter; \ie $\xi \lesssim 0.93$.
The plots in Fig.~\ref{gauge2} illustrated the fact that the upper bounds on the \pp parameter from the tensor perturbations and the bispectrum are weaker in 
comparison to the generated scalar perturbations.

In this paper we also discussed the gauge quanta production during inflation where quanta are produced by a gravitational coupling between the inflaton and
the gauge field. In this case also the leading constraint on the \pp parameter comes from the non-production of long-lived PBHs. In this model, the bound on
$\xi$ from the tensor bispectrum was stringent than the tensor-to-scalar ratio (see Table~\ref{tab1}). 

In the second scenario where the inflaton field coupled to a scalar field, the model was free of DM PBHs overproduction in the CMB observational range if the amplitude of the generated bump in the scalar power spectrum, $A_i$ is less than $4\times10^{-4}$. This bound is weaker than the one in \cite{0909.0751}, however, our result was completely model independent.

As we mentioned, in this paper we did not consider any specific potential for the inflaton and the coupled field. It would be interesting to consider different
inflationary potentials and study the \pp in these models. In this case one can put bound on the parameter(s) of the model such as the coupling 
constant via the non-production of PBHs. We leave this subject for a future work. It is also worth mentioning that in all studied models the produced PBHs can not give the right relic abundance of DM according to the observational results.

\section*{Acknowledgments}

EE is grateful to the hospitality of the ICTP-SAIFR, S\~{a}o Paulo during her visit under the TWAS 2015 Fellowship for Research and Advanced Training.
She is indebted to Eiichiro Komatsu for motivating the importance of this work during the MIAPP workshop ``Cosmology after Planck''. 
EE would like to thank Hassan Firouzjahi, Rafael Porto and Ivonne Zavala for their useful comments on the draft.

\end{document}